\documentclass[aip,jap,reprint,showpacs,showkeys,superscriptaddress,nofootinbib]{revtex4-1}
\usepackage[usenames,dvipsnames]{xcolor}
\usepackage[colorlinks=true, linktocpage=true, linkcolor=blue, citecolor=gray]{hyperref}
\usepackage[utf8]{inputenc}
\usepackage{graphicx}
\usepackage{amsfonts}
\usepackage{amssymb}
\usepackage{amsmath}
\usepackage{color}
\usepackage{ulem}
\usepackage{xspace}

\begin{document}

\title{Determination of the blocking temperature of  magnetic nanoparticles:\\ The good, the bad and the ugly }

\author{P. \surname{Mendoza Zélis}}
\email{pmendoza@fisica.unlp.edu.ar}
\affiliation{IFLP-CCT- La Plata-CONICET and Departamento de Física, Facultad de Ciencias Exactas, C. C. 67, Universidad Nacional de La Plata, 1900 La Plata, Argentina}
\affiliation{Departamento de Ciencias Básicas, Facultad de Ingeniería, Universidad Nacional de La Plata, 1900 La Plata, Argentina}

\author{I.J. Bruvera}
\email{bruvera@fisica.unlp.edu.ar}
\affiliation{IFLP-CCT- La Plata-CONICET and Departamento de Física, Facultad de Ciencias Exactas, C. C. 67, Universidad Nacional de La Plata, 1900 La Plata, Argentina}

\author{M.Pilar Calatayud}
\affiliation{Aragon Institute of Nanoscience (INA), University of Zaragoza, 50018, Spain}
\affiliation{Condensed Matter Physics Department, Science Faculty, University of Zaragoza, 50009, Spain}

\author{G.F. Goya}
\affiliation{Aragon Institute of Nanoscience (INA), University of Zaragoza, 50018, Spain}
\affiliation{Condensed Matter Physics Department, Science Faculty, University of Zaragoza, 50009, Spain}

\author{F.H. S\'anchez}
\affiliation{IFLP-CCT- La Plata-CONICET and Departamento de Física, Facultad de Ciencias Exactas, C. C. 67, Universidad Nacional de La Plata, 1900 La Plata, Argentina}

\begin{abstract}
In a magnetization vs. temperature (M vs. T) experiment, the blocking region of a magnetic nanoparticle (MNP) assembly is the interval of T values were the system begins to respond to an applied magnetic field H when heating the sample from the lower reachable temperature. The location of this region is determined by the anisotropy energy barrier depending on the applied field H, the volume V, the magnetic anisotropy constant K of the MNPs and the observing time of the technique. In the general case of a polysized sample, a representative blocking temperature value $T_B$ can be estimated from  ZFC-FC experiments as a way to determine the effective anisotropy constant.

In this work, a numerical solved Stoner-Wolfharth two level model with thermal agitation is used to simulate ZFC-FC curves of monosized and polysized samples and to determine the best method for obtaining a representative $T_B$ value of polysized samples. The results corroborate a technique based on the T derivative of the difference between ZFC and FC curves proposed by Micha \textit{et al} (the good) and demonstrate its relation with two alternative methods: the ZFC maximum (the bad) and inflection point (the ugly). The derivative method is then applied to experimental data, obtaining the $T_B$ distribution of a polysized $Fe_3O_4$ MNP sample suspended in hexane with an excellent agreement with TEM characterization.
\end{abstract}
\maketitle

\section{Introduction}

\begin{figure}[h!]
  \centering
    \includegraphics[width=0.3\textwidth]{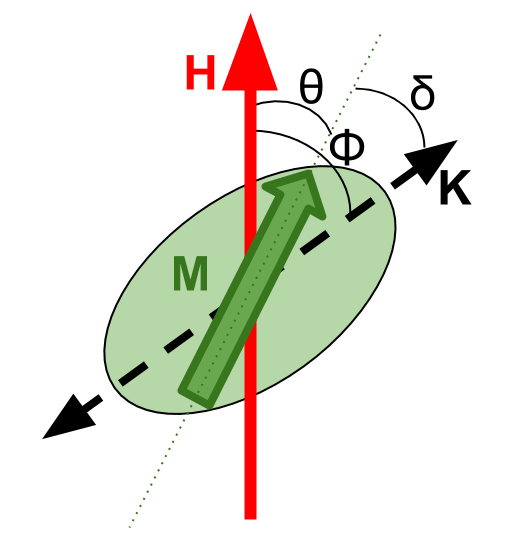}

\caption{\label{NPM}MNP model. The energy is determined by the angle $\delta$ between the magnetization $M$ and the field $H$, and the angle $\theta$ between $M$ and the easy axis $K$. For calculation simplicity the angle $\phi= \theta+\delta$ between $K$ and $H$ is used. }
\end{figure}

Magnetic nanoparticles (MNPs) are been extensively studied due to their multiple applications in technology \cite{Reiss2005} and biomedicine \cite{pankhurst2003applications, pankhurst2009progress}. Particles with sizes in the range $[5 ,100] nm$ \cite{rosensweig2002heating} present a magnetic behaviour determined by its volume, shape and composition, matrix viscosity and temperature, among other factors.
In the simplest (however very useful) model, the MNPs of volume $V$ and saturation magnetization $M_s$ are considered as almost spherical ellipsoids with a permanent moment $m=M_sV$ and a preferential magnetization axis (easy axis) in which the anisotropy energy $E_K=KV \sin^2[\delta]$ is minimum, being $K$ the effective anisotropy density constant and $\delta$ the angle between $m$ and the easy axis. If the MNPs are fixed in the matrix and separated one from each other by a distance $d>3V^{1/3}$, dipolar interactions can be neglected\cite{woinska2013magnetic} and the energy of the system can be expressed as the sum of the anisotropy energy and the Zeeman energy $E_H=-mH\cos[\theta]$:

\begin{equation}
E=E_K+E_H,
\label{energ}
\end{equation}
with $\theta$ the angle between $m$ and $H$ (fig. \ref{NPM}). This configuration is usually called Stoner-Wolfharth system since the publication of a work\cite{stoner1948mechanism} in which the authors perform a numerical calculation of  the $M$ vs. $H$ curves of ordered systems with different orientations \textit{i.e.} systems of identical MNPs with a single value of $\phi$, and the $M$ vs.$ H$ curve of a disordered system \textit{i.e.} with a uniform distribution of $\phi$ values. Since no thermal agitation was considered by Stoner and Wolfharth, their calculations were made just finding the positions $\theta_i$ of the minima of equation \ref{energ} for each value of $H$.

In order to calculate the temperature dependence of the magnetic response for MNPs systems, it is necessary to consider the effect of thermal fluctuations that allow transitions between stable configurations. Doing so, it is possible to simulate $M$ vs. $T$ experiments as the extensively performed Zero Field Cooling-Field Cooling (ZFC-FC) routine. In this kind of experiments, a sample  is cooled from a temperature where all particles show superparamagnetic behaviour to the lowest reachable temperature (usually around $3 K$), then, a small constant field usually lower than $8 kA/m$ is applied, and the sample is heated to a temperature high enough to observe an initial growth and subsequent decrease in magnetization, i.e. were the sample show again superparamagnetic behaviour .The sample is then cooled again to the lowest temperature with the constant field still applied.

In the ideal case of a monosized, non interacting MNPs sample; a narrow temperature region should exist in which the system performs a transition between irreversible and reversible regimes. When heating with applied field, the thermal energy $kT$ is initially much smaller than the anisotropy barrier $KV$ so the magnetization remains null. Due to the exponential dependence of the Néel relaxation time with temperature\cite{bean1959superparamagnetism}, when $kT \sim KV$, the magnetization grows rapidly until its thermodynamic equilibrium value, defining the aforementioned transition region. The Blocking Temperature $T_B$ can be considered as the inflection point of this growing and its experimental determination is an important goal of the MNPs characterization.

Real samples always present a size dispersion, usually reasonably well described by a log-normal distribution. Different particle size implies different anisotropy barrier $KV$ and therefore a different $T_B$ for each size fraction, so, in real ZFC-FC experiments, the blocking region is wide and the representative $T_B$ value is not well defined. There are several different criteria used to define a representative $T_B$ from ZFC-FC data of polysized samples. Some authors maintain the inflexion point (IP) criterion\cite{schmitz2015x} while others report the maximum ZFC magnetization temperature (MAX)\cite{de2015particle}\cite{sankar2000magnetic}, being all this criteria still in discussion\cite{tournus2011magnetic}. In an alternative approach, Micha \textit{et al.}\cite{micha2004estimation} propose a method in which the $T_B$ distribution is obtained from the T derivative of the difference between ZFC and FC curves. An approximated theoretical justification for this method was presented by Mamiya \textit{et al} \cite{mamiya2005extraction}.

In this work, a SW model with thermal agitation is applied to obtain the temporal dependence of the magnetization $M(t)$ for an ordered system of identical MNPs in a similar way to previous works of Lu\cite{lu1994field}, Usov\cite{usov2009hysteresis} and Carrey\cite{carrey2011simple}. Temperature dependence $dM(T)/dT$ is then obtained in order to numerically simulate the ZFC-FC curves. In contrast to the method implemented by Usov\cite{usov2011numerical} where a stair-step approximation for the time evolution of the temperature was used, we consider a continuous time evolution. Finally, an ordered polysize system response is simulated by linear combination of the monosize curves weighted by a discrete log-normal distribution.

The validity of the method proposed by Micha \textit{et al} was verified by comparing the $T$ derivative of  this ZFC-FC curve with the $T_B$ distribution obtained from the inflection points of each volume of the distribution. The resultant mean blocking temperature value $<T_B>$ is then compared, for several volume distributions, with the commonly used criteria for a representative $T_B$: the inflection point temperature IP and the maximum MAX of the ZFC curve.

Additionally, Micha’s method is tested with experimental data of a magnetite MNPs frozen ferrofluid suspended in hexane comparing the obtained $T_B$ distribution with the one calculated from the TEM size information. In order to obtain an ordered system, the ferrofluid was frozen while a large constant was field applied.

\section{Model}
A SW-like model with thermal agitation and zero width energy minima approximation was developed in order to obtain ZFC-FC curves of fixed MNPs with size dispersion. Only the simplest case of an ordered system was considered, with all the MNPs oriented (easy axis orientation) in the direction of the field. This situation can be achieved experimentally by freezing a ferrofluid sample under a sufficiently strong applied field ($\sim 7T$).

\subsection{Magnetization vs. time equation}
For a system of identical, fixed, non interacting MNPs of volume $V$, anisotropy constant $K$ and saturation magnetization $M_s$, with their anisotropy axes parallel to an external field $H$, the energy can be expressed as the sum of the anisotropy energy $E_k$ and the Zeeman energy $E_h$\cite{stoner1948mechanism}:

\begin{equation}
\begin{split}
E=E_k+E_h&=KV Sin(\theta)^2- \mu_0 M_s V Cos(\theta)\\
&=KV \bigg(Sin(\theta)^2-2h Cos(\theta)\bigg),
\end{split}
\end{equation}
being $h=H/H_k$ and $H_k=\mu_0 M_s/2K$.

\begin{figure}
  \centering
    \includegraphics[width=0.5\textwidth]{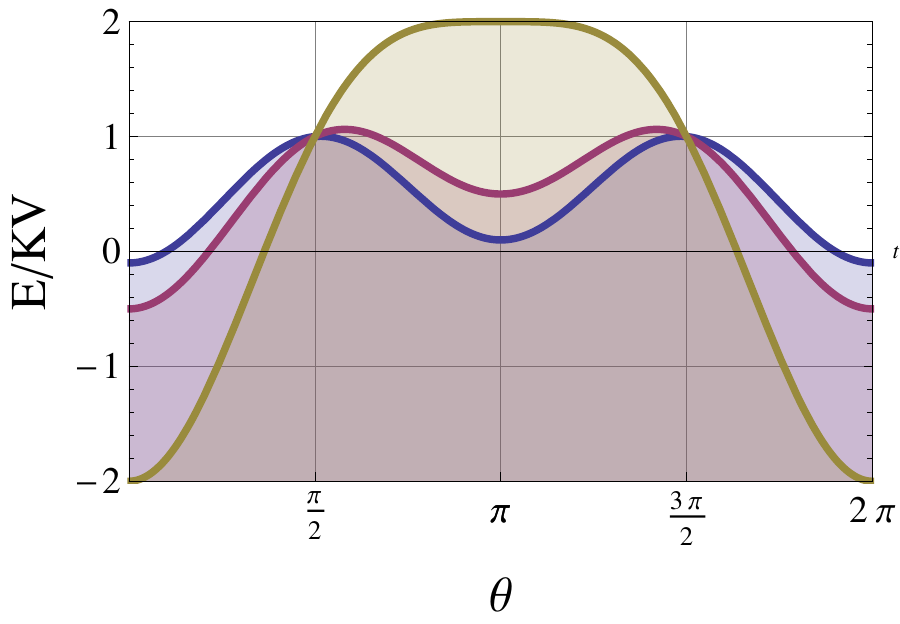}
\caption{\label{landscapes}Energy landscapes as a function of theta for a fixed MNP under a magnetic field at $\theta=0$. Each color stands for a different value of the reduced field factor $2h$. Blue: $2h=0.1$, purple: $2h=0.5$, gold: $2h=2$.}
\end{figure}

In the range $\theta=[0, 2\pi]$, this energy landscape presents two minima, of $E(0)=-2KV_h$ and $E(\pi)=2KV_h$ and a maximum of $E(\arccos(-h))= KV\bigg(1+h^2\bigg) $(fig. \ref{landscapes}).

The frequency of thermal inversions between minima $i$ and $j$ is the inverse of the Nèel relaxation time\cite{balian2006microphysics}\cite{neel1949},:
 \begin{equation}
 f_{ij}=f_0 e^{D_{ij}/(kT)}
 \label{frec}
 \end{equation}
with $f_0$ the ``intrinsic frequency'', times the Boltzman ``success probability'' depending on the ratio between thermal energy $kT$ and barrier height $D_{ij}$. The barrier between minima is symmetric for $h=0$ with $D_{du}=D_{ud}=KV$ (naming $u$ and $d$ to $\theta=0$ and $\theta=\pi$ directions respectively) and smaller for inversion to the field direction otherwise:

\begin{equation}
\begin{aligned}
\Delta _{ud}=\ E(\arccos(-h))-E(\pi)=KV\bigg(1+h\bigg)^2 \\
\Delta _{du}=\ E(\arccos(-h))-E(0)=KV\bigg(1-h\bigg)^2
\end{aligned}
\end{equation}

It is a good approximation to consider the same $f_0$ value for both frequencies\cite{aharoni2000introduction}.

Sample magnetization $M$ in the direction of the applied field can be expressed in terms of saturation magnetization $M_s$  and the number of particles per unit volume magnetized in each direction $N_u$ and $N_d$:
\begin{equation}
 M=\bigg(N_u - N_d\bigg) M_s / N =M_s \bigg(2N_u / N - 1\bigg),
\end{equation}
with $N$ the total number of particles per unit volume. So the time derivative of the magnetization can be written in terms of the population variation which is equal to the actual population times the inversion probability to each direction

\begin{equation}
\dfrac{d M}{d t}= 2\dfrac{M_s}{N} \dfrac{dN_u}{dt}.\label{eq:deriv}
\end{equation}

\begin{equation}
\dfrac{dN_u}{dt}= -\dfrac{dN_d}{dt}= f_{d \rightarrow u}N_d- f_{u \rightarrow d} N_u.
\label{dNdt}
\end{equation}
so the time derivative of the relative magnetization m is determined by the transcendental equation

\begin{equation}
\dfrac{d m}{d t}= 2 f_0 e^{-C\big(1+h^2\big)}\lbrace \sinh(2C\,h)- m\,\cosh(2C\,h)\rbrace.
\label{dmdt}
\end{equation}
where $C=KV/kT$.

\subsection{Magnetization vs. Temperature equation: ZFC-FC simulation}
Temperature dependence of the magnetization can be obtained from \ref{dmdt} via the equation

\begin{equation}
\dfrac{dm}{dT}=\dfrac{d m}{d t}\dfrac{d t}{d T}.
\label{der}
\end{equation}

For a linear temperature variation $T(t)=B t+T_0$, the magnetization derivative is

\begin{equation}
\dfrac{dm}{dT}= \dfrac{1}{B}\dfrac{d m}{d t} =\dfrac{2 f_0}{B} e^{-C\big(1+h^2\big)}\lbrace \sinh(2C\,h)- m\,\cosh(2C\,h)\rbrace
\label{dmdT}
\end{equation}

\begin{figure}
  \centering
    \includegraphics[width=0.5\textwidth]{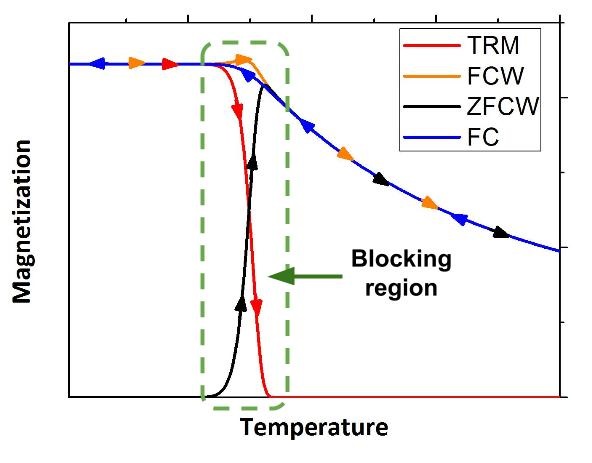}

\caption{\label{mono}Simulation result for an ordered assembly of MNPs. The system is first cooled with zero field applied from a high temperature where all particles show superparamagnetic behaviour (ZFC, no showed), then,  the field is turned on and the system is heated  beyond the blocking region (ZFCW). Maintaining the applied field, the system is cooled (FC). The final heating can be performed with (FCW) or without applied field (TRM).}
\end{figure}

Solving this equation by numerical methods it is possible to simulate a ZFC-FC experiment for a monosize sample. A Matlab script based on the ODE15s\cite{shampine1997matlab} function was developed.
An example of the result for a monosize assembly of ordered MNPs is shown in figure \ref{mono}. Line colours stand for different parts of the routine.

During the warming after zero field cooling (ZFCW for this chapter, usually called just ZFC), the exponential dependence of the inversion frequency with temperature in equation \ref{frec} determines a narrow ``blocking region'' wherein the MNPs, which were ``blocked'' at low temperature begin to respond to the field. Magnetization grows with temperature since the applied field has decreased the energy barrier for $\theta=\pi$ to $\theta=0$ inversion. The magnetization increasing reverts when thermal energy is much higher than the barrier, so the difference between inversion frequencies in each direction tends to disappear. The blocking temperature $T_B$ of the system is then defined as the inflection point of the magnetization growing when heating.

When the system is cooled again (FC), magnetization grows monotonically while the barrier height difference between inversions becomes increasingly significant against thermal energy. This growth stops when thermal energy becomes too low for inversions to occur within the experimental window time. If the system is then heated maintaining the applied field (FCW), magnetization values are the same than FC except for the blocking region where there is a small increase due to the assembly getting closer to the equilibrium state. If the final warming is done with no applied field (Thermal Remanent Magnetism, TRM), magnetization drops to zero in the blocking region when thermal energy is enough for the wells populations to equilibrate.

The magnetization values $M_p$ for a polysized sample are obtained by linear addition of the $M_{Vi}$ values for each contemplated size $V_i$, weighted by the corresponding volume and log-normal distribution $LnN(V_i)$ value:

\begin{equation}
M_p(T)=\dfrac{\sum^N_{i=1}M_{Vi}(T)V_i LnN(V)}{\sum^N_{i=1}V_i LnN(V)},
\label{suma}
\end{equation}

The $V_iLnN(V_i)$ product stands for the relative volume distribution.

\begin{figure}
  \centering
    \includegraphics[width=0.5\textwidth]{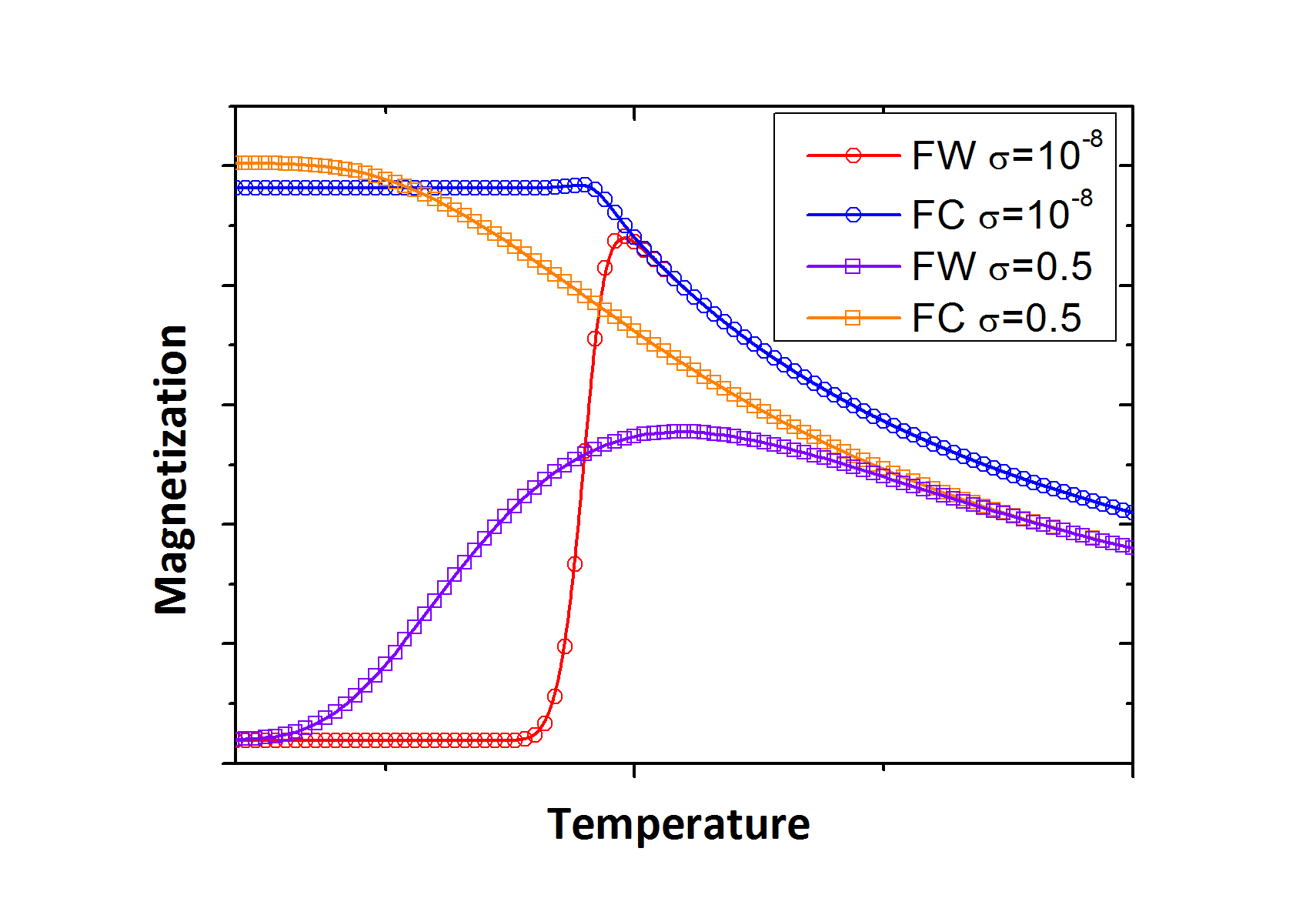}

\caption{\label{comp}Comparison between ZFC-FC simulations for small ($\sigma=10^{-8}$) and large ($\sigma=0.5$) dispersion systems.}
\end{figure}

Figure \ref{comp} shows the comparison between ZFC-FC simulations for samples with different size dispersion expressed as the scale parameter $\sigma$ of the log-normal number distribution. A much wider transition region can be seen for the bigger dispersion so the different aforementioned criteria would define very separated values for a representative $T_B$.

\section{Blocking temperature determination}
\subsection{Micha's method verification}
\begin{figure}
  \centering
    \includegraphics[width=0.5\textwidth]{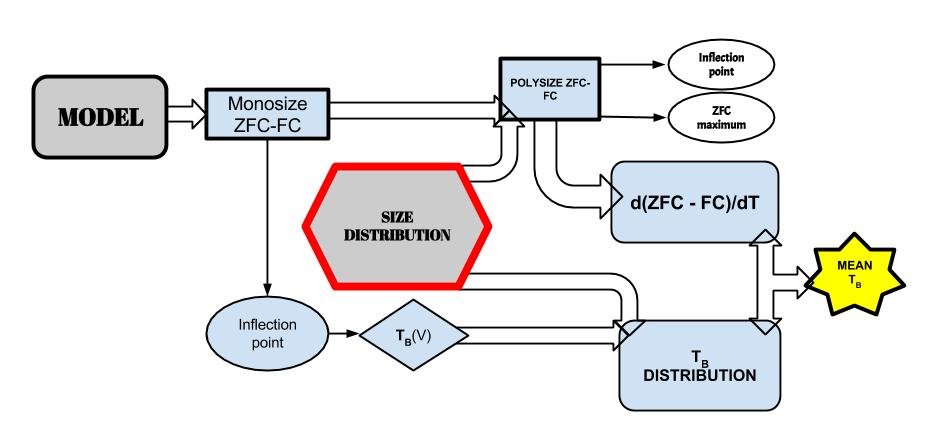}

\caption{\label{esq}Scheme of the method verification. Monosize ZFC-FC curves are simulated from the dM(T)/dT equations of the model. In one path, $T_B$ for every particle size is calculated as the IP of the monosize ZFC curve. In the other path, a polysize ZFC-FC curve is simulated by linear addition of the monosize values. Then the T derivative of the difference ZFC-FC is calculated.
}
\end{figure}

In order to verify Micha’s method, several polysize ZFC-FC experiments were simulated using differents parameter sets varying $\sigma$ and the mean radius. For each one of the used sets, the T derivative of the ZFC-FC difference was calculated. Then, the $T_B$ distribution was obtained from the monosize curves that were added to construct the polyzise simulation in equation \ref{suma}: a ZFC curve was calculated for each class of the size distribution so each $T_B$ class  comes from a volume class, maintaining the same relative height. Also IP and MAX values of the polysize ZFC curve were calculated and compared with the mean value $<T_B>$ of the distribution in each simulation (fig. \ref{esq}).

\begin{figure}
  \centering
    \includegraphics[width=0.5\textwidth]{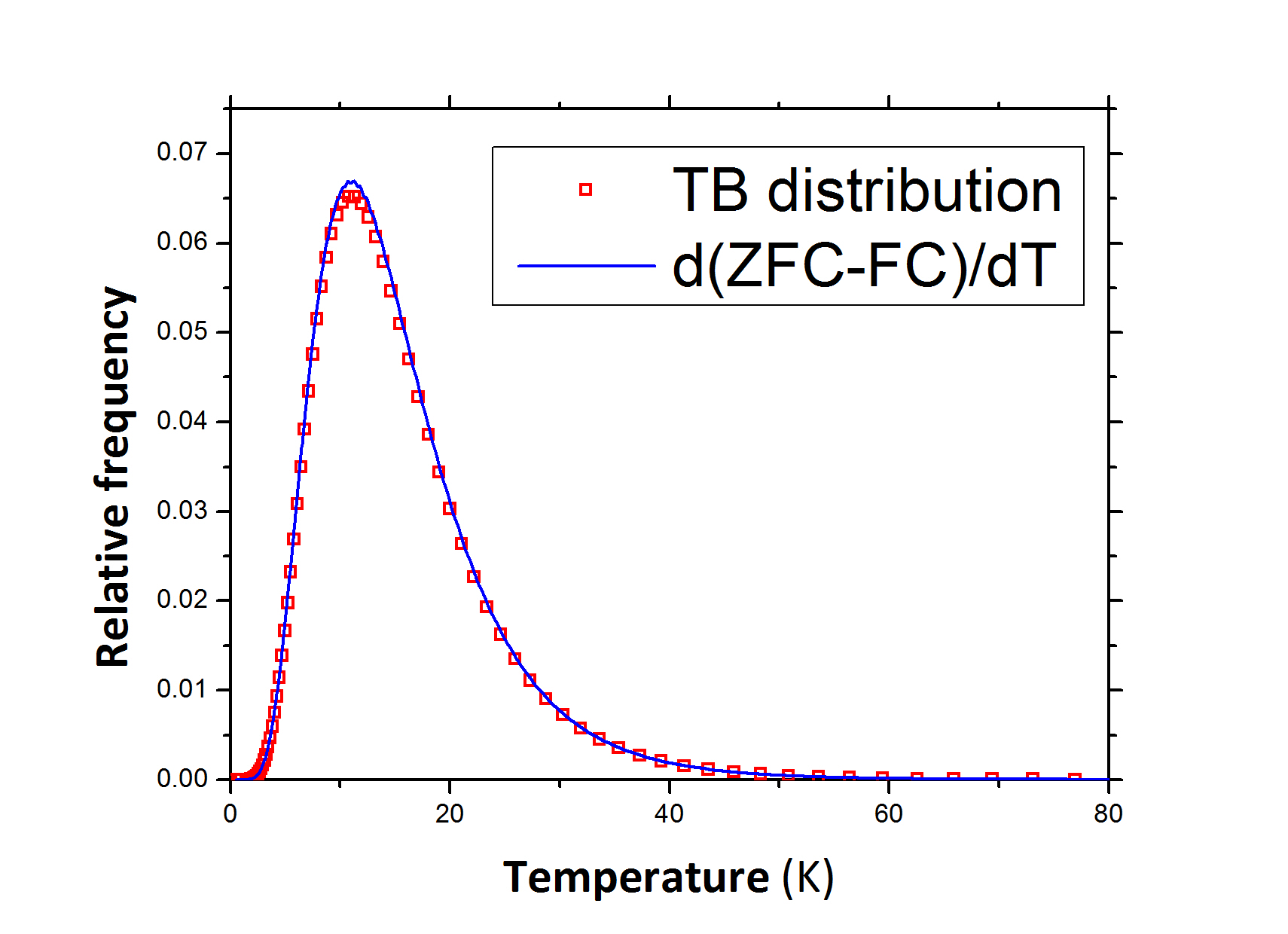}

\caption{\label{dist}Comparison between the $T_B$ distribution obtained directly from the size distribution used in the simulation and the derivative d(ZFC-FC)/dT of the simulated curves.}
\end{figure}

In all cases, the $T_B$ distribution and the ZFC-FC derivative are identical. Figure \ref{dist} shows the results for the simulation with 4.5 nm mean radius, $\sigma=0.5, 16 kJ/m^3$ anisotropy constant and a $4 K/min$ heating rate.

\begin{figure}
  \centering
    \includegraphics[width=0.5\textwidth]{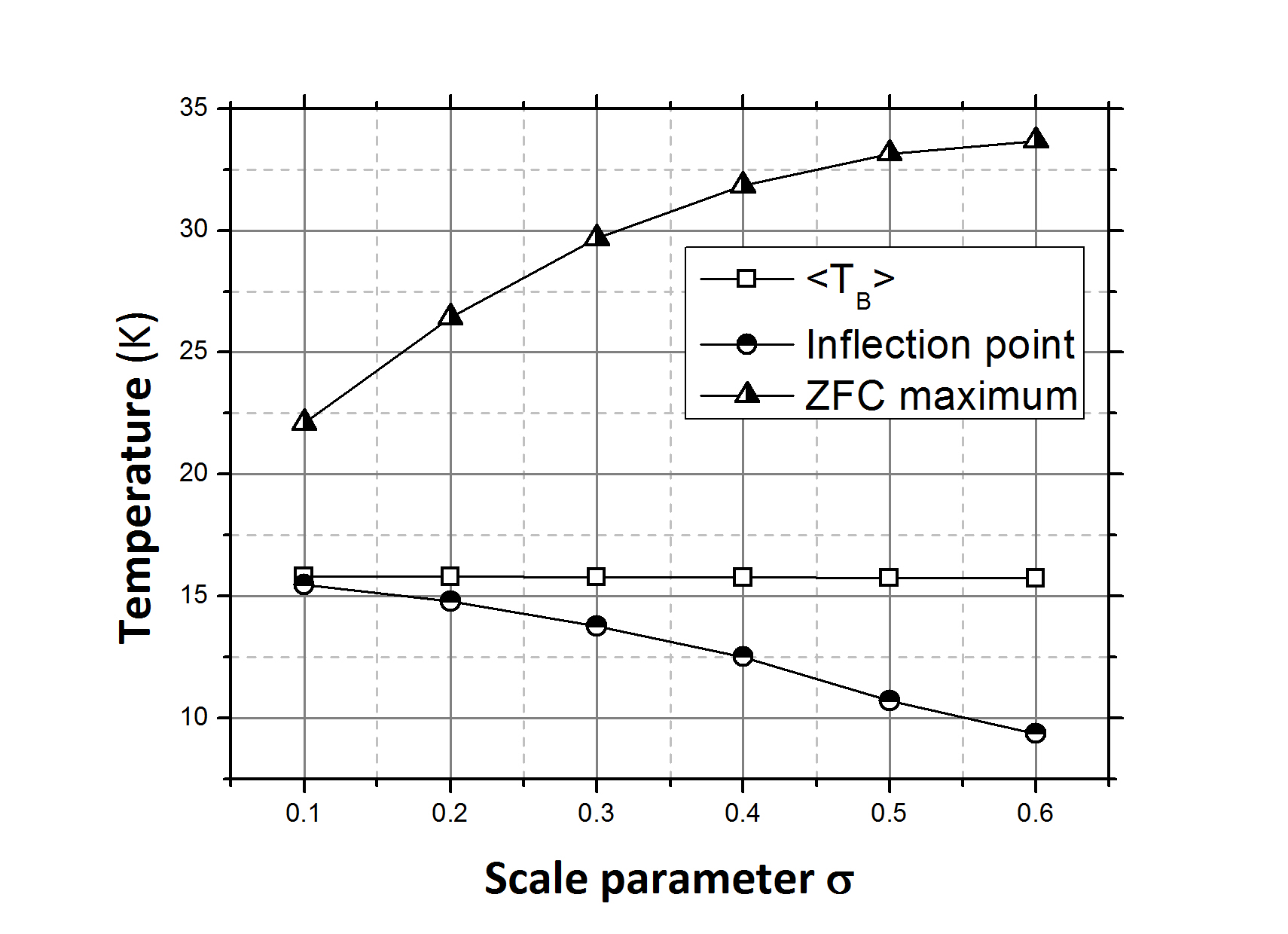}

\caption{\label{med}Values of $T_B$ mean, MAX and IP of the simulated curves as a function of the scale parameter $\sigma$ for $4.5 nm$ mean radius, $16 kJ/m^3$ anisotropy constant and $4 K/min$ heating rate. }
\end{figure}

Also, for a set of  ZFC-FC curves calculated with the same mean volume, saturation magnetization, heating rate and anisotropy constant, by increasing scale parameter $\sigma$, $<T_B>$ stays constant while the polysize curve IP  shifts to smaller temperatures and MAX shifts in opposite direction. Figure  \ref{med} shows the results for $4.5 nm$ mean radius, $16 kJ/m^3$ anisotropy constant, $4 K/min$ heating rate and  $\sigma=[0.1, 0.6]$.

\begin{figure}
  \centering
    \includegraphics[width=0.5\textwidth]{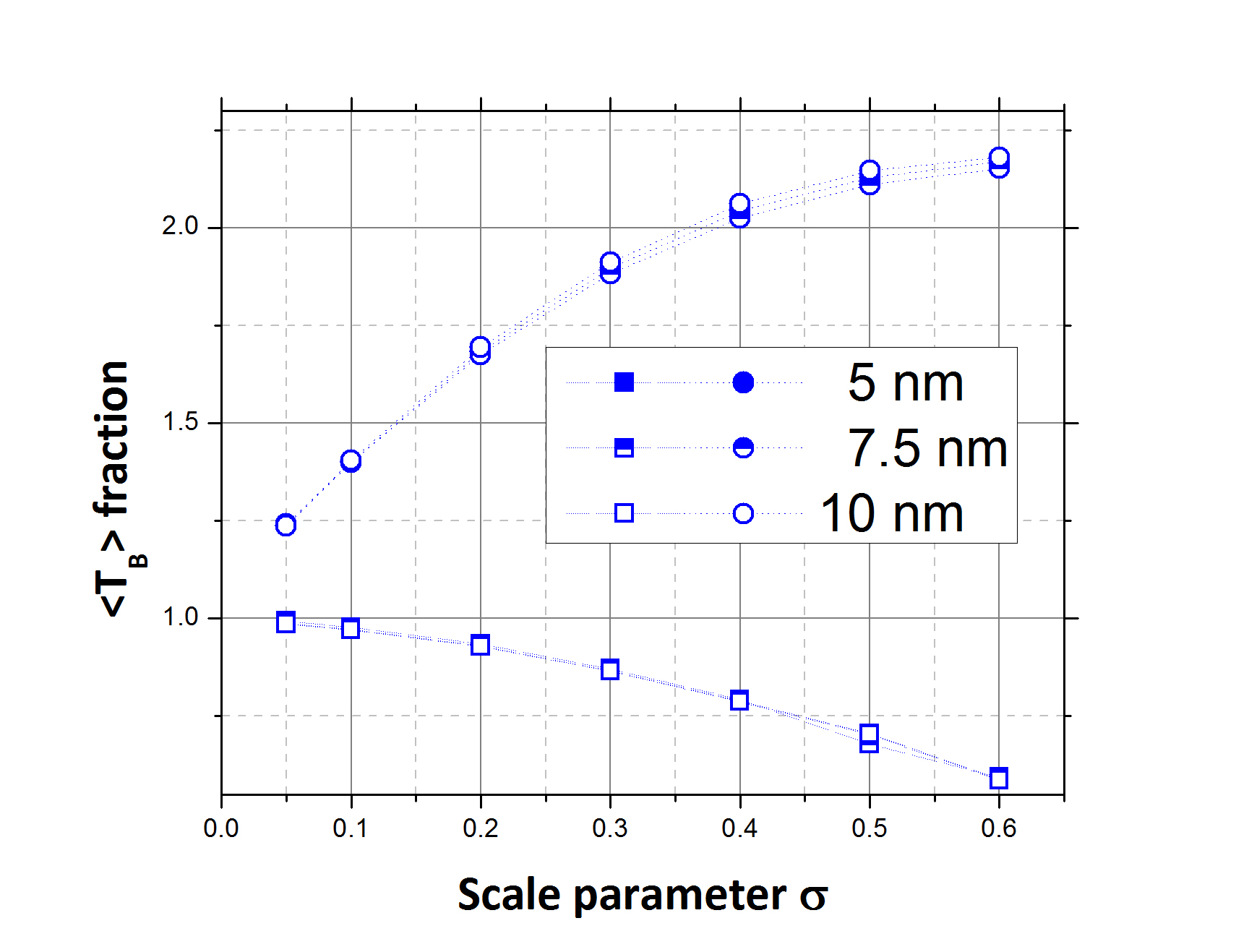}

\caption{\label{norm}Values of MAX (circles) and IP (squares) of the simulated curves divided by $T_B$ mean for different MNP radii.  The behavior is the same for all sizes.}
\end{figure}

This behaviour is the same in the hole studied size range. By normalizing IP values by the $T_B$ mean, all points fall in the same curve as shown in figure \ref{norm} while the variation for MAX is small.

\begin{figure}
  \centering
    \includegraphics[width=0.5\textwidth]{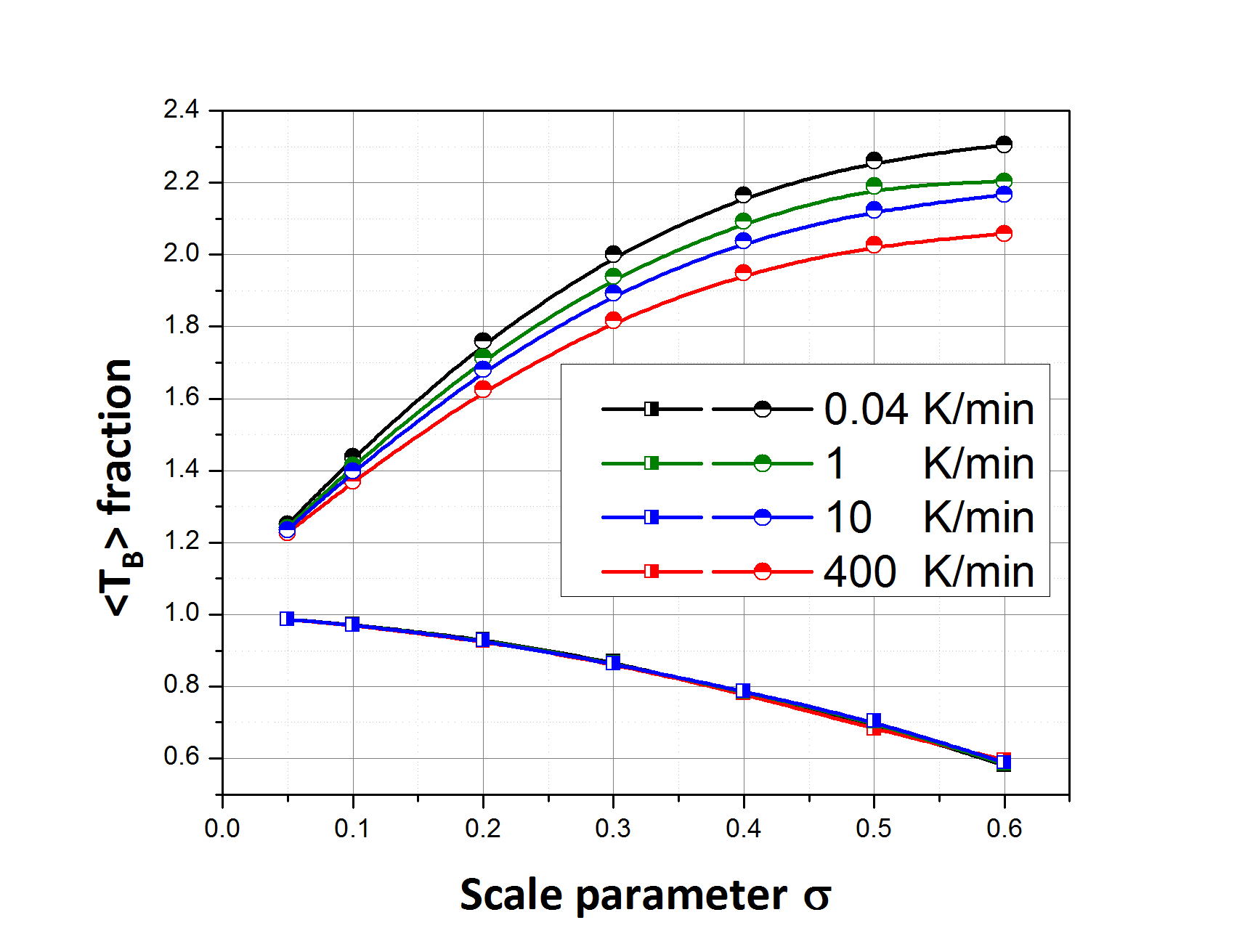}

\caption{\label{VsRate}MAX (circles) and IP (squares) relative to $<T_B>$  for $10 nm$ NPM mean radius at different temperature rates.}
\end{figure}

Varying the heating rate and $K$ does not affect the relation IP/$<T_B>$. Meanwhile, the MAX/$<T_B>$ ratio changes strongly in the range $[0.04; 400] K/min$ and noticeably in the range $[1; 10] K/min$ and also depends on the $K$ value. Figure \ref{VsRate} shows the results of varying the heating rate for $R_m=10 nm$, $K=16 kJ/m^3$ and $M_s=281 kA/m$.

\begin{figure}
  \centering
    \includegraphics[width=0.5\textwidth]{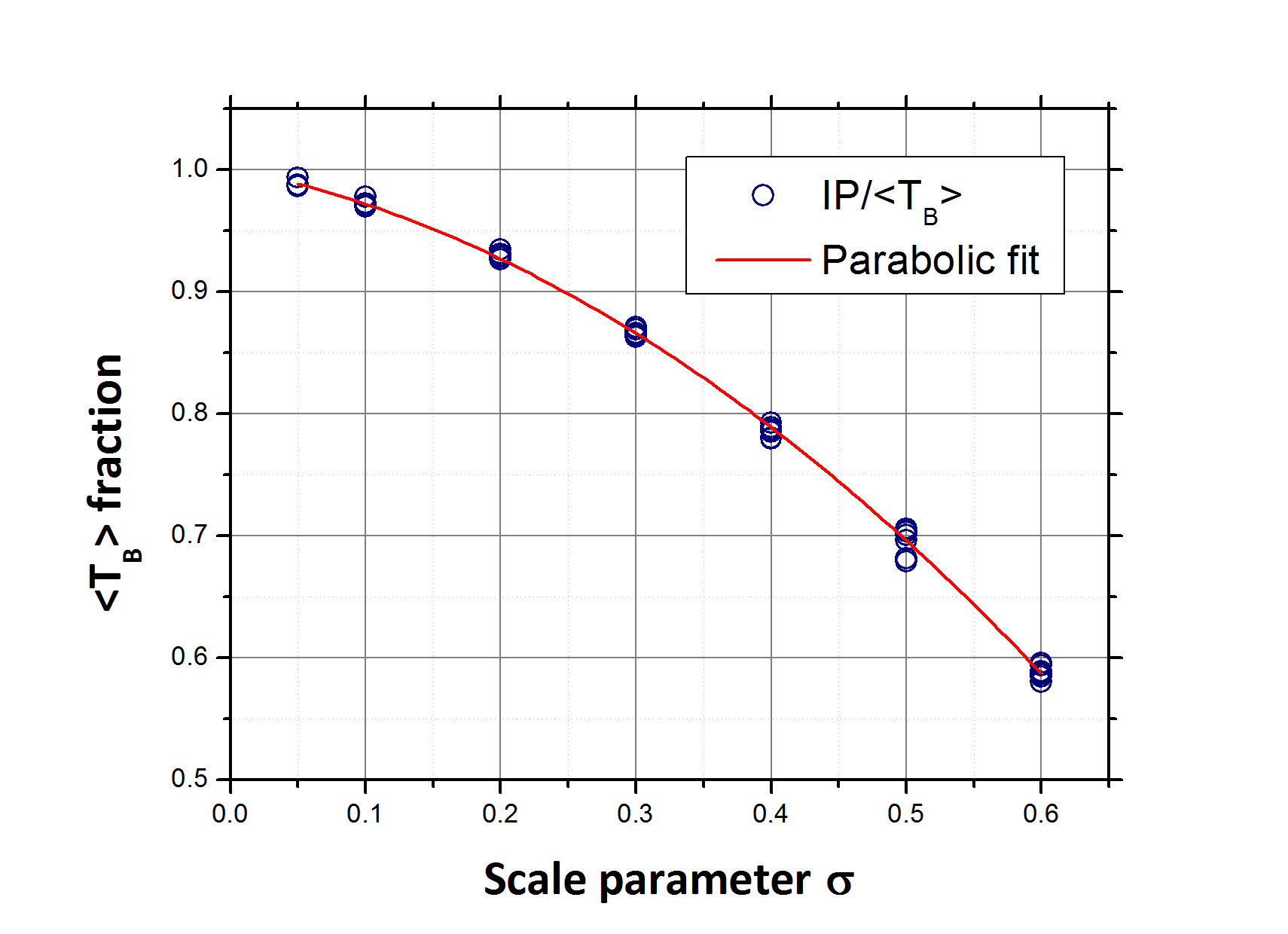}

\caption{\label{aj}Parabolic fitting of the universal curve IP/$<T_B>$.}
\end{figure}

Figure \ref{aj} shows a parabolic fit over the $IP/<T_B>$ values obtained for all the simulations. The curve is universal with small fluctuations due to numeric resolution. The obtained polynomial with fitting errors is $\frac{IP}{<T_B>}(\sigma)=1.00(2)-0.21(2)\sigma-0.79(2)\sigma^2$.

\subsection{Experimental application}
\begin{figure}
  \centering
    \includegraphics[width=0.5\textwidth]{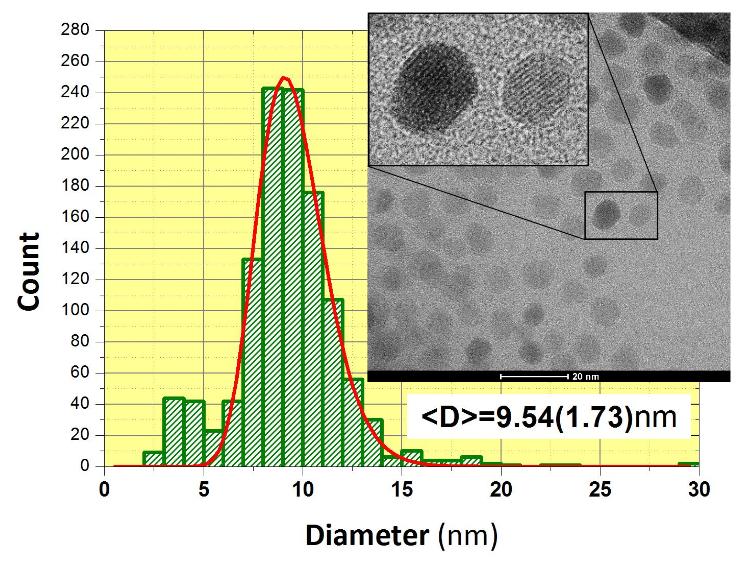}

\caption{\label{TEM}Size distribution from TEM images. Inset: TEM image example with a magnification showing the crystallinity of the particles.}
\end{figure}

The Micha’s analysis was conducted on ZFC-FC measurements of a FF of magnetite MNPs suspended in hexane with a concentration of $12(1) g/L$.
TEM images were taken in order to determine the size distribution of the particles (fig \ref{TEM}). A narrow log-normal number diameter distribution ($LnN(x)$) was obtained with a $9.54 nm$ mean and a $1.73 nm$ standard deviation. The relative TEM volume distribution was obtained from this results and fitted with a $xLnN(x)$ function obtaining a scale parameter $\sigma=0.55(2)$.

\begin{figure}
  \centering
    \includegraphics[width=0.5\textwidth]{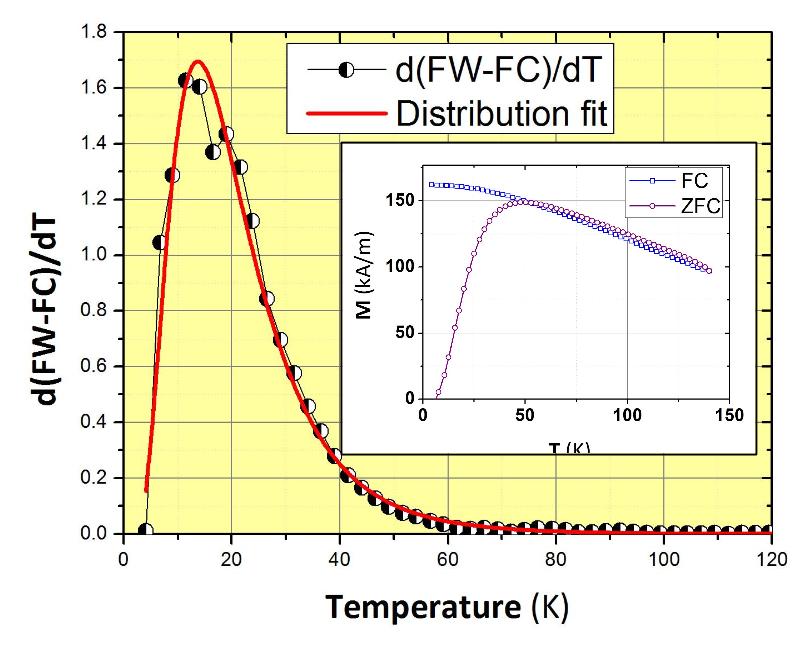}

\caption{\label{deri}Log-normal fit on the d(ZFC-FC)/dT derivative. Inset: ZFC and FC experimental curves.}
\end{figure}

The ZFC-FC routine was carried with a $2.4 K/min$ rate and a $8 kA/m$ field on an encapsulated FF sample frozen under a $7 T$ field in order to obtain an ordered system with all the MNP easy axes oriented parallel to the field.
The ZFC-FC derivative was calculated and fitted with a $xLnN(x)$ distribution using the TEM $\sigma$ as a fixed parameter with a very good correspondence (figure \ref{deri}).

\begin{figure}
  \centering
    \includegraphics[width=0.5\textwidth]{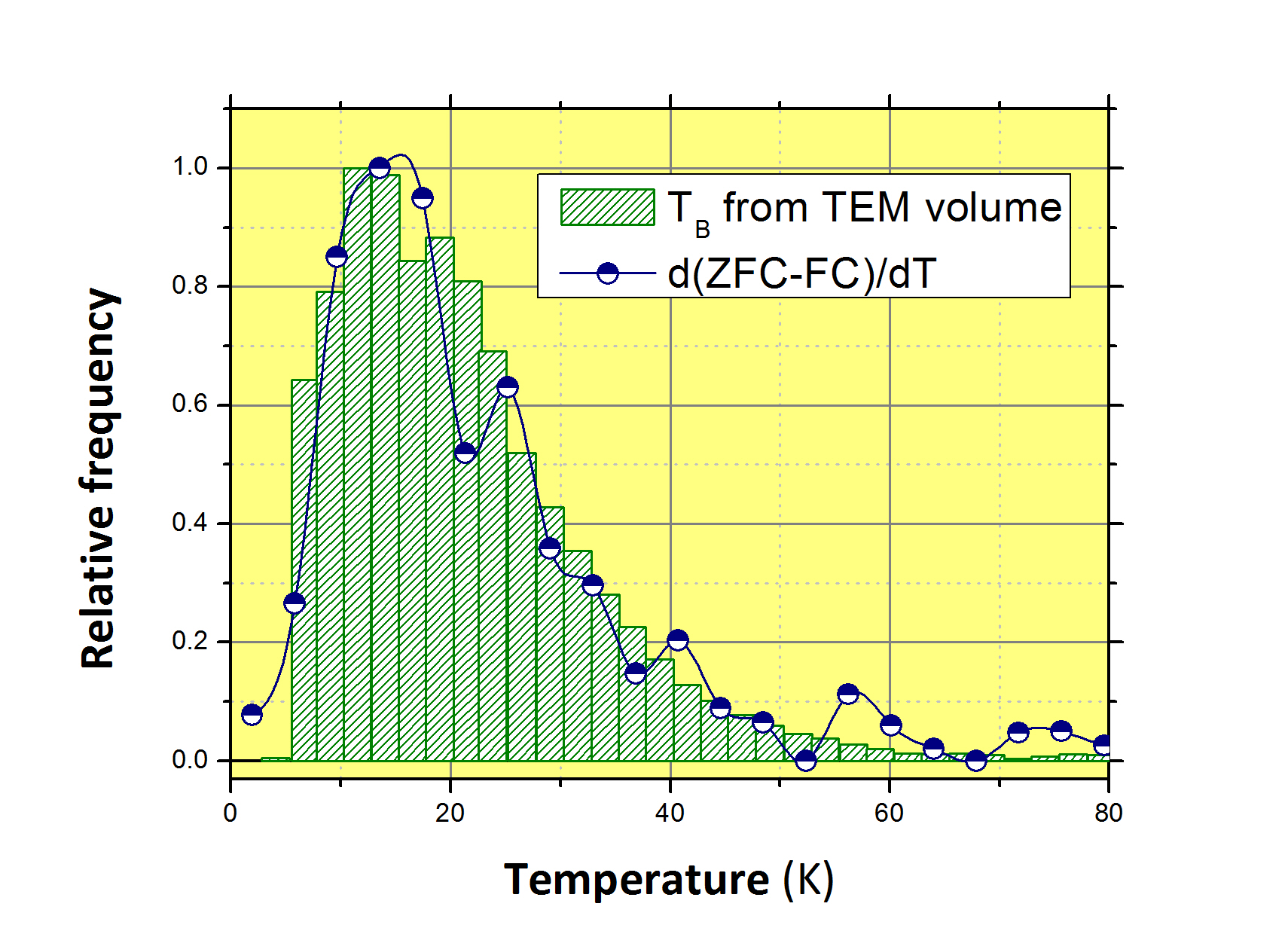}

\caption{\label{vs} d(ZFC-FC)/dT derivative together with $T_B$ distribution from TEM volume obtained by fitting $K$ for maximum coincidence.}
\end{figure}

Figure \ref{vs} shows the comparison between $T_B$ distribution obtained from TEM information and the ZFC-FC derivative curve. The translation from TEM volume to $T_B$ was made considering the blocking condition in which the inversion time of the MNPs is approximately equal to the measurement time of the magnetization value:
\begin{equation*}
\tau(K,V,h,T ) = \tau_0 \exp\bigg(\frac{KV}{kT_B}(1-h)^2\bigg)\approx \tau_m
\end{equation*}
\begin{equation}
\Rightarrow T_B=\frac{KV(1-h)^2}{k \log(\tau_m/\tau_0)},
\label{tau}
\end{equation}

where $\tau_0=1/f_0$ is the inverse of the intrinsic inversion frequency. For a known volume distribution, this comparison can be used to determine the effective $K$ value as the one that maximizes the coincidence between TEM and ZFC-FC distributions. In this case, a value of $34(2) kJ/m^3$ was obtained with a very good correspondence between TEM and ZFC-FC data. This calculation implies some approximations: $M_s$ is considered independent from the temperature in the region of interest, and the relaxation time expression used for the blocking condition \ref{tau} considerate only the inversions in the direction of the field. While the first approximation is very reasonably, the blocking condition expression is accurate only in experiments with high $\mu H/(kT)$ ratios, where the reversal frequency are much smaller for the inversions to the antiparallel state.

Additionally, the IP/$<T_B>$ ratio was calculated obtaining a value of $0.7(1)$, compatible with polynomial expression obtained from the simulations.

\section{Discussion and conclusions}

The validity of the Micha’s method to determine the $T_B$ distribution of non interacting MNPs assembly was demonstrated by numerical simulations and experimental data analysis.

A Stoner-Wolfarth model with thermal agitation was developed in order to simulate the ZFC-FC curves of polysized MNPs assembles. From this simulation it was clearly demonstrated that the temperature derivative of the ZFC-FC difference is in full coincidence with the $T_B$ distribution of the sample, calculated as the inflection points of each size ZFC curve.
Additionally, it came clear from the results that the maximum (MAX) and the inflection point (IP) of the polysized ZFC curve are affected not only by the mean size of the particles, but by the size dispersion. Thus neither IP or MAX are direct estimators of the mean $T_B$. This is an interesting result since these values are commonly used in magnetic characterizations and can lead to estimate $T_B$ values far from the mean. As an example, for a sample with $\sigma=0.5$ and a heating rate of $4 K/min$, IP$=0.7<T_B>$ and MAX$=2.12<T_B>$.
Nevertheless, it was found that the IP$/<T_B>$ ratio depends exclusively on $\sigma$, while MAX$/<T_B>$ changes with K and the heating-cooling rate. This behavior reveals a connection between IP, one of the most commonly used $T_B$ criteria and the actual mean value of the blocking temperature. For a sample with known $\sigma$, the mean blocking temperature could be obtained from the universal curve presented in this work. This approach of obtaining size distribution information from an universal curve was presented before by Hansen \textit{et al} \cite{hansen1999estimation} using a rougher model.

In the present development status, the ZFC-FC simulation algorithm does not include a ``measurement time'' parameter. Just the heating rate is used, so the authors assume that the simulated data points represent the values for ``instantaneous'' measurements. Therefore, the magnetization values obtained from the simulation won't be equivalent to the ones obtained in an experiment with the same parameters.

The Micha’s method was then applied to characterize a sample of magnetite nanoparticles coated with oleic acid and suspended in hexane.
The volume distribution of the sample was obtained from TEM analysis showing a narrow log-normal shape with a mean diameter of $9.54 nm$ and a standard deviation of $1.73 nm$.
In order to obtain an ordered system, the ferrofluid was frozen under a $7 T$ magnetic field. Then, a ZFC-FC routine was carried and a $T_B$ distribution was obtained from the data. This distribution was fitted with a $xLnN[x]$ function using the scale parameter $\sigma$ obtained from the TEM data as a fixed fitting parameter. The high goodness of the fitting supports the validity of Micha’s method and the low influence of magnetic interaction between particles which is consistent with the particle to particle distance imposed by the FF concentration. Additionally, the resultant IP/$<T_B>$ values is consistent with the universal curve obtained from the simulations.

Finally, the effective anisotropy constant of the particles was estimated as the value which gives the maximum coincidence between the ZFC-FC $T_B$ distribution and the one obtained from the TEM volume.

The results obtained in this work constitute just a first example of the potential of the presented model in combination with experimental characterization. There is work in progress to enhance the simulation algorithm in order to include the measurement time as a parameter and to considerate both inversions processes in the blocking condition.

\bibliography{biblioHT}

\begin{thebibliography}{10}

\bibitem{Reiss2005}
G\"{u}nter Reiss and Andreas H\"{u}tten.
\newblock Magnetic nanoparticles: Applications beyond data storage.
\newblock {\em Nature Materials}, 4(10):725--726, oct 2005.

\bibitem{pankhurst2003applications}
Quentin~A Pankhurst, J~Connolly, SK~Jones, and JJ~Dobson.
\newblock Applications of magnetic nanoparticles in biomedicine.
\newblock {\em Journal of physics D: Applied physics}, 36(13):R167, 2003.

\bibitem{pankhurst2009progress}
QA~Pankhurst, NTK Thanh, SK~Jones, and J~Dobson.
\newblock Progress in applications of magnetic nanoparticles in biomedicine.
\newblock {\em Journal of Physics D: Applied Physics}, 42(22):224001, 2009.

\bibitem{rosensweig2002heating}
Ronald~E Rosensweig.
\newblock Heating magnetic fluid with alternating magnetic field.
\newblock {\em Journal of magnetism and magnetic materials}, 252:370--374,
  2002.

\bibitem{woinska2013magnetic}
Magdalena Woi{\'n}ska, Jacek Szczytko, Andrzej Majhofer, Jacek Gosk, Konrad
  Dziatkowski, and Andrzej Twardowski.
\newblock Magnetic interactions in an ensemble of cubic nanoparticles: A monte
  carlo study.
\newblock {\em Physical Review B}, 88(14):144421, 2013.

\bibitem{stoner1948mechanism}
Edmund~C Stoner and EP~Wohlfarth.
\newblock A mechanism of magnetic hysteresis in heterogeneous alloys.
\newblock {\em Philosophical Transactions of the Royal Society of London A:
  Mathematical, Physical and Engineering Sciences}, 240(826):599--642, 1948.

\bibitem{bean1959superparamagnetism}
CP~Bean and JD~Livingston.
\newblock Superparamagnetism.
\newblock {\em Journal of Applied Physics}, 30(4):S120--S129, 1959.

\bibitem{schmitz2015x}
Carolin Schmitz-Antoniak.
\newblock X-ray absorption spectroscopy on magnetic nanoscale systems for
  modern applications.
\newblock {\em Reports on Progress in Physics}, 78(6):062501, 2015.

\bibitem{de2015particle}
Patricia de~la Presa, Yurena Luengo, Victor Velasco, MP~Morales, M~Iglesias,
  Sabino Veintemillas-Verdaguer, Patricia Crespo, and Antonio Hernando.
\newblock Particle interactions in liquid magnetic colloids by zero field
  cooled measurements: Effects on heating efficiency.
\newblock {\em The Journal of Physical Chemistry C}, 119(20):11022--11030,
  2015.

\bibitem{sankar2000magnetic}
Sandra Sankar, AE~Berkowitz, D~Dender, JA~Borchers, RW~Erwin, SR~Kline, and
  David~J Smith.
\newblock Magnetic correlations in non-percolated co--sio 2 granular films.
\newblock {\em Journal of magnetism and magnetic materials}, 221(1):1--9, 2000.

\bibitem{tournus2011magnetic}
F~Tournus and A~Tamion.
\newblock Magnetic susceptibility curves of a nanoparticle assembly ii.
  simulation and analysis of zfc/fc curves in the case of a magnetic anisotropy
  energy distribution.
\newblock {\em Journal of Magnetism and Magnetic Materials}, 323(9):1118--1127,
  2011.

\bibitem{micha2004estimation}
JS~Micha, B~Dieny, JR~R{\'e}gnard, JF~Jacquot, and J~Sort.
\newblock Estimation of the co nanoparticles size by magnetic measurements in
  co/sio 2 discontinuous multilayers.
\newblock {\em Journal of Magnetism and Magnetic Materials}, 272:E967--E968,
  2004.

\bibitem{mamiya2005extraction}
H~Mamiya, M~Ohnuma, I~Nakatani, and T~Furubayashim.
\newblock Extraction of blocking temperature distribution from
  zero-field-cooled and field-cooled magnetization curves.
\newblock {\em IEEE transactions on magnetics}, 41(10):3394--3396, 2005.

\bibitem{lu1994field}
Jing~Ju Lu, Huei~Li Huang, and Ivo Klik.
\newblock Field orientations and sweep rate effects on magnetic switching of
  stoner--wohlfarth particles.
\newblock {\em Journal of applied physics}, 76(3):1726--1732, 1994.

\bibitem{usov2009hysteresis}
NA~Usov and Yu~B Grebenshchikov.
\newblock Hysteresis loops of an assembly of superparamagnetic nanoparticles
  with uniaxial anisotropy.
\newblock {\em Journal of Applied Physics}, 106(2):023917, 2009.

\bibitem{carrey2011simple}
Julian Carrey, Boubker Mehdaoui, and Marc Respaud.
\newblock Simple models for dynamic hysteresis loop calculations of magnetic
  single-domain nanoparticles: Application to magnetic hyperthermia
  optimization.
\newblock {\em Journal of Applied Physics}, 109(8):083921, 2011.

\bibitem{usov2011numerical}
NA~Usov.
\newblock Numerical simulation of field-cooled and zero field-cooled processes
  for assembly of superparamagnetic nanoparticles with uniaxial anisotropy.
\newblock {\em Journal of Applied Physics}, 109(2):023913, 2011.

\bibitem{balian2006microphysics}
Roger Balian and Dirk Haar.
\newblock {\em From microphysics to macrophysics: methods and applications of
  statistical physics}, volume~2.
\newblock Springer Science \& Business Media, 2006.

\bibitem{neel1949}
L~Neel.
\newblock Influence des fluctuations thermiques a l aimantation des particules
  ferromagnetiques.
\newblock {\em Comptes Rendus de l'Académie des Sciences}, 228:664--668, 1949.

\bibitem{aharoni2000introduction}
Amikam Aharoni.
\newblock {\em Introduction to the Theory of Ferromagnetism}, volume 109.
\newblock Oxford University Press, 2000.

\bibitem{shampine1997matlab}
Lawrence~F Shampine and Mark~W Reichelt.
\newblock The matlab ode suite.
\newblock {\em SIAM journal on scientific computing}, 18(1):1--22, 1997.

\bibitem{hansen1999estimation}
Mikkel~Fougt Hansen and Steen M{\o}rup.
\newblock Estimation of blocking temperatures from zfc/fc curves.
\newblock {\em Journal of Magnetism and Magnetic Materials}, 203(1):214--216,
  1999.

\end{thebibliography}

\end{document}